# Real Space Imaging of Field-Driven Decision-Making in Nanomagnetic Galton Boards

H. Arava[1], D. Sanz-Hernandez[2], J. Grollier[2], and A. K. Petford-Long[1]

[1]*Materials Sciences Division (MSD), Argonne National Laboratory, Argonne, IL 60439 USA*
[2]*Unité Mixte de Physique, CNRS, Thales, Université Paris-Saclay, 91120 Palaiseau, France*

**Abstract**

A possible spintronic route to hardware implementation for decision making involves injecting a domain wall into a bifurcated magnetic nanostrip resembling a Y-shaped junction. A decision is made when the domain wall chooses a particular path through the bifurcation. Recently, it was shown that a structure like a nanomagnetic Galton Board, which is essentially an array of interconnected Y-shaped junctions, produces outcomes that are stochastic and therefore relevant to artificial neural networks. However, the exact mechanism leading to the robust nature of randomness is unknown. Here, we directly image the decision-making process in nanomagnetic Galton Boards using Lorentz transmission electron microscopy. We identify that the stochasticity in nanomagnetic Galton Boards arises as a culmination of: (1) topology of the injected domain wall, (2) local disorder, and (3) strength of the applied field. Our results pave the way to a detailed understanding of stochasticity in nanomagnetic networks.

**Introduction**

With the world's energy needs for computing expected to surpass the global energy output in the next few decades, it is essential that alternative low power solutions to traditional computing are found.[1] A solution to the energy problem is to use the spin of an electron for memory and computing applications, i.e., spintronics based on nanomagnetism.[2] Novel spintronic computing architectures that move away from the traditional von Neumann architectures offer an energy-efficient future for computing.[3,4] Essential to many of the novel computing architectures, such as artificial neural networks and neuromorphic computing, is the reliance on devices that can reliably create stochasticity. Stochasticity in nanomagnetism is often achieved by one or more of the following methods: (1) thermal/field fluctuations,[5-7] (2) disorder,[8] and (3) tuning of the energy landscape.[9,10] Components generating stochasticity can be incorporated directly into computing architectures by providing probabilistic logic,[11] assigning weights to a decision-making process,[12] generating random numbers, and other novel purposes. In addition, reliable stochasticity has uses in information theory, such as encryption and related national security applications.

Recently, a nanomagnetic version of the classical Galton Board was experimentally shown to be robust at producing tunable stochasticity.[13] In a classical Galton board, a series of pegs are positioned at locations that mimic the geometry of Pascal's triangle. Metal balls dropped from the top of such a board undergo a series of decision-making events in terms of how they pass each peg, before dropping off into bins at the bottom. A nanomagnetic realization of such a Galton Board is made of interconnected bars of Permalloy arranged onto a honeycomb lattice, with the lattice terminated with elongated bars that serve as the equivalent of the classical bins. A domain wall is injected into the top of the nanomagnetic Galton board, passes through the bars causing magnetization reversal that is field driven, and terminates as a reversed magnetization state in one or more of the elongated bars at the bottom of the lattice, which can then be read out. Even though robustness of the stochasticity in nanomagnetic Galton Boards has previously been established, the mechanism leading to emergence of stochasticity has not been probed experimentally. In order to incorporate structures such as Galton Boards into computing architectures and other novel applications, it is essential that the physics of the domain wall motion leading to stochasticity is fully understood.

Earlier reports on domain wall motion in Y-shaped nanostructures and hexagonal artificial spin ice[14] (ASI) lattices provide us with glimpses into the different mechanisms of domain wall motion in bifurcated structures. Of note is the topological explanation to path selectivity in which vortex domain walls of particular chirality (and polarity) predominantly favor one path over another.[15] This chiral path selectivity for vortex wall motion arises from the winding number conservation in the Y-shaped structures, due to the inherent topological nature of the vortex wall as the structure carries with it defects with winding number equal to $-\frac{1}{2}$ along its diagonal, while retaining a defect with winding number equal to $+1$ at the center. Other experiments involving hexagonal arrays of connected nanomagnets, such as ASIs, provide us with statistical understanding of domain wall motion in multiple bifurcated paths. For example, it was found that the chiral nature of the domain walls leads to non-random propagation of domain walls in a Permalloy honeycomb lattice.[14] It should be noted that the path selectivity of chiral domain walls is not always robust and is sensitive to local disorder.[16] Additionally, path selectivity can be tuned by changing the bifurcation geometry: in an asymmetric Y-shaped junction it is possible to enforce that domain wall motion always follows a pre-determined path.[17] Domain wall motion along bifurcated paths is interesting both from a fundamental physics point of view and because of the possibility of implementing Y-shaped junctions in applications. The robustness of stochasticity in nanomagnetic Galton Boards provides an excellent opportunity to study domain wall motion in bifurcated paths while also targeting their possible use in computing applications. Using Lorentz Transmission Electron Microscopy (LTEM) on a nanomagnetic Galton Board, we are able to carry out high resolution imaging of local changes to the magnetic microstructure that enable us to identify the exact mechanisms leading to the emergence of domain wall motion stochasticity in Y-shaped junctions.

In this work, we directly image domain wall motion in nanomagnetic Galton Boards with two geometries, using LTEM. The difference between them lies in the number of injection pads, with one structure containing one injection pad (in-focus TEM image shown in Fig. 1) and the other having four injection pads (not shown here). The structures are fabricated using electron beam lithography from Permalloy with a thickness of 25 nm, and each of the bars of the honeycomb lattice has a length of 1 μm and a width of 200 nm. The structures were deposited onto TEM grids coated with electron-transparent SiN membranes. It should be noted that the lower two edges of the vertices at the tops of the bars in the honeycomb lattice contains curved edges to help with domain wall depinning, with these curved vertices being ~150% larger in area than those at the bottoms ends of the bars (the two vertex types are labeled as v1 and v2 in Fig. 1). The experiments were carried out by first applying a global field of ~ 75 Oe, parallel to the injection pad, as indicated in Fig. 1. Once the structures were fully magnetized, we implemented a stepped-field protocol in the opposite direction to the initial saturation. As the field was increased, with a step size of ~0.2 Oe, the domain wall motion was imaged using a JEOL 2100F TEM dedicated for-high resolution Lorentz microscopy. In the Fresnel mode of LTEM, the magnetic contrast is imaged in out-of-focus images, with bright and dark contrast corresponding to the local magnetic domain wall structures. All LTEM images displayed in the manuscript were recorded in the under-focus condition, with the bright and dark regions corresponding to domain walls: bright circular regions indicate magnetic vortices (e.g.

vortex domain walls) that have a clockwise (CW) rotation of magnetization (chirality), while the darker regions have counterclockwise (CCW) magnetization. We observed that the domain wall motion begins at the location of the injection pad and subsequently terminates at the bottom of the lattice. The observed domain wall motion was similar in both the 1-pad and the 4-pad Galton boards. The experiments were repeated three times in 1-pad and 4-pad Galton boards, and additional evidence for domain wall motion was collected in 40 smaller structures each resembling a Y shaped junction. We found that the decision regarding which path to follow through the nanomagnetic Galton Boards occurs at three distinct lattice locations (see Fig. 1): (1) at the injection pads, (2) via domain wall interactions in the vertices, and (3) via changes to the local magnetic microstructure along the bars. We find that the complex interplay between random domain wall chirality at creation in the pads, chirality-driven path selectivity at the vertices, and reversal of chirality at the vertices and in the bars explains the emergence of tunable stochasticity.

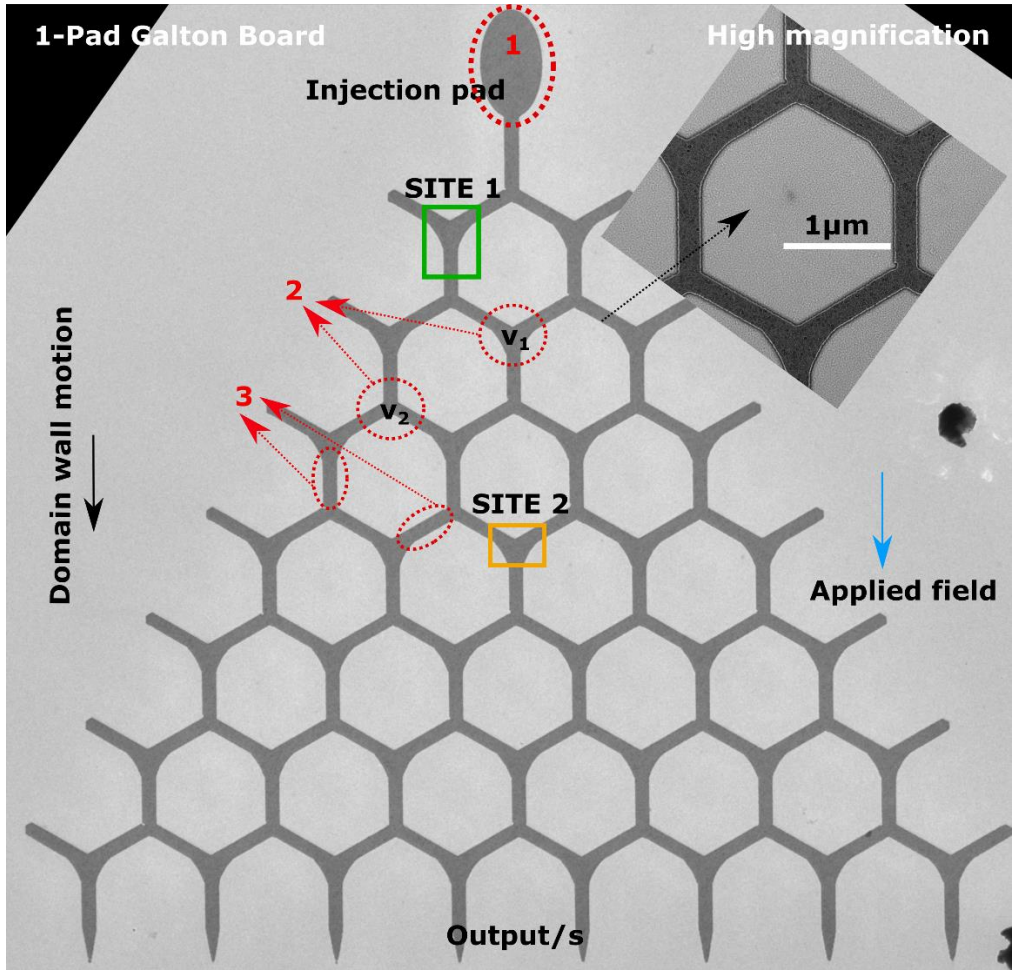

**Fig. 1**. In-focus TEM image of 1-pad nanomagnetic Galton Board. Domain wall is injected by the pad at the top. Upon field reversal, the domain wall proceeds to the bottom of the structure leading to a change in magnetization in one or more of the outputs. The three locations that affect decision making are labeled 1, 2, and 3. Each unit cell of the honeycomb consists of two types of vertices, v1 and v2, with v1 larger in area than v2. In the inset is a high magnification image of the nanomagnetic Galton Board, indicating a uniform structural morphology of the prepared sample.

**Chirality-driven path selectivity:** In Fig. 2a-f, we show the path selectivity of an injected vortex domain wall as it encounters the first vertex in the honeycomb lattice. We found that domain walls with CCW chirality, such as the bright regions seen in Figs. 2a and 2c, predominantly took a left turn (*y*-branch), as indicated in Figs. 2b and 2d. On the other hand, walls with CW chirality, such as the dark region in Fig. 2e, took a right turn (*x*-branch), as shown in Fig. 2f. Such a chirality-dependent path selectivity is likely due to topological conservation of winding numbers at the Y-shaped junctions.[15] For clarity, we have included a schematic of the observed magnetization directions in each bar represented by blue and red arrows corresponding to the magnetization along the direction of the applied field or opposite to it.

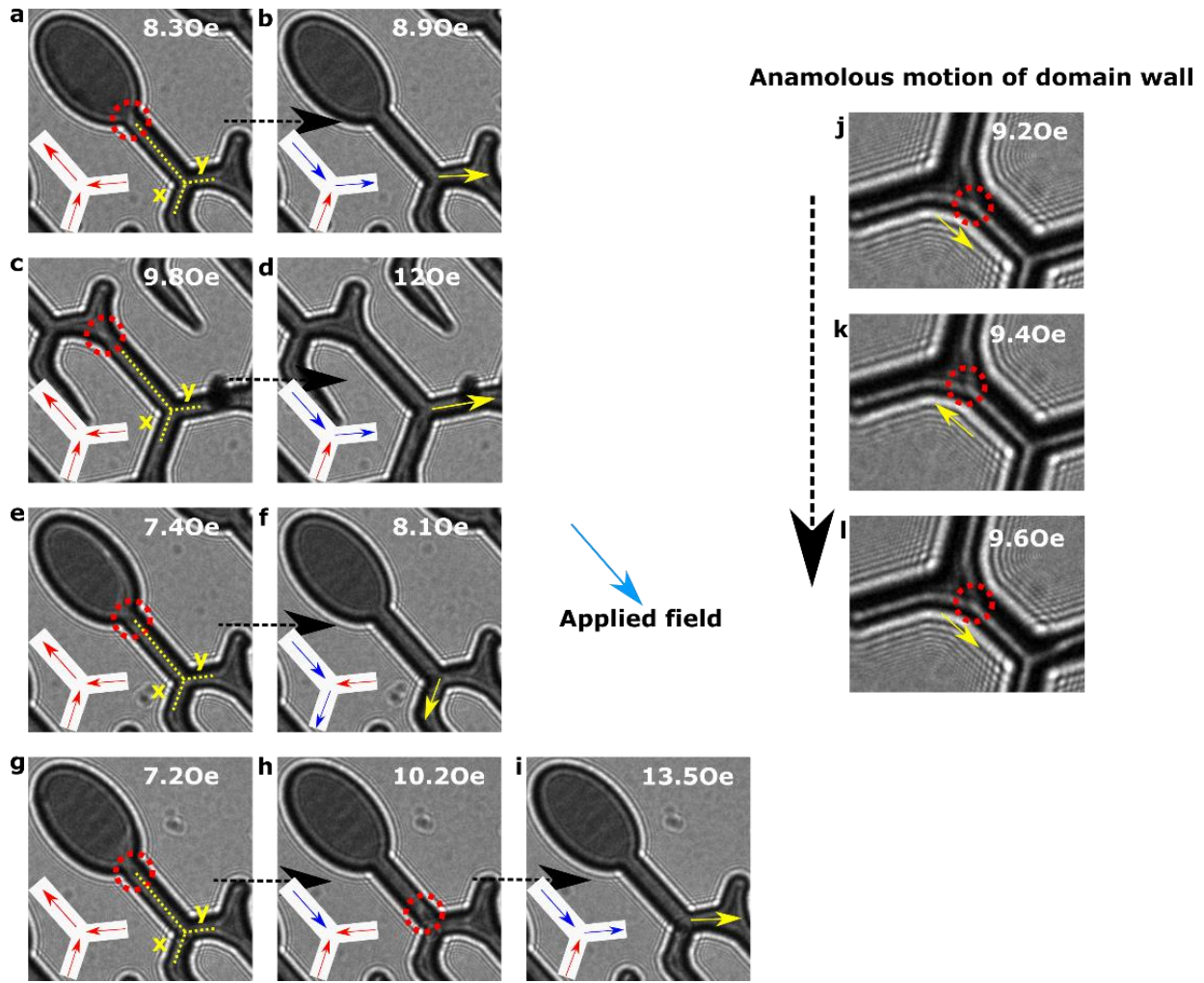

**Fig. 2**. Path selectivity of the injected domain walls. (a-f) Injected domain walls (ringed in red) prefer x or y branches at each vertex dependent on their chirality (dark or bright wall contrast). The switching of the domain wall can be inferred by comparing the width of the dark Fresnel fringe contrast on either side of the nanostructure, with the wider dark regions switching sides as the magnetization is reversed. (g-i) A vortex wall likely undergoes a change to its chirality, leading to a loss of path selectivity. (j-l) An anomalous motion of a domain wall in which the domain wall moves in the direction opposite to the applied field. It is possible this could be related to pinning effects or even Walker breakdown. The magnetization orientation in the bars is represented by red and dark blue arrows.

**Chirality transformations within straight segments (bars):** We observed that the domain walls may undergo a change to their chirality or internal magnetic structure, as they travel along one of the bars in the honeycomb lattice. Fig. 2g shows an LTEM image of a CW domain wall at an applied field of 7.2 Oe, that undergoes a transformation into a complex magnetic texture, shown in Fig. 2h, at an applied field of 10.2 Oe. The large difference of 3 Oe between the events observed in Fig. 2g and Fig. 2h indicate that the domain wall seen in Fig. 2g might have been subject to pinning or to dynamic transformations, such as Walker breakdown.[18] Walker breakdown is evidenced by a domain wall undergoing a precessional motion, i.e., the domain wall would have instances of motion in the opposite direction to the applied field. In our field-driven experiments, we have observed some evidence for this behavior in the form of an anomalous response of a domain wall to the applied field, as shown in Figs. 2j-l. We therefore believe that the morphological and chirality changes to the domain walls during their motion along the elongated bars is due to both local disorder resulting from microstructure and lithographic defects, and to effects related to the strength of applied field such as pinning and Walker breakdown. It should be noted that the Walker breakdown field ($H_w$) in our structures is calculated to be ~11.5 Oe based on the model described elsewhere[19] for Py bars that are 200 nm wide and 25 nm thick, as used in the experiment. The model does not account for pinning and other microstructure related effects.

The initial injection of the domain wall and the motion at the first bifurcation, which typically occurred at fields below 10 Oe (i.e. < $H_w$), are likely not influenced by Walker breakdown effects. We have not visually observed any precessional motion in the domain walls at the first bifurcation, within the timescales we are able to resolve using the microscope. We should therefore note that the initial chirality-dependent path selectivity of the domain walls, which is a topological effect based on conservation of the local winding number. We have also observed instances where the domain wall is pinned at a particular location, as for example in Fig. 2g in which a CCW vortex domain wall is pinned at the vertex, and then depinned after an increase of 3.3. Oe in the applied field, see Fig. 2i. Local pinning of domain walls leads to a loss of path selectivity, as shown in Figs. 2g-i, where a likely CCW vortex domain wall (Fig. 2g) underwent a reversal in its chirality to a CW vortex domain wall (Fig. 2h). The change in chirality can be inferred in the subtle change of the contrast associated with the domain wall (ringed in red, from Fig. 2g to Fig. 2h), which would explain the domain wall motion along the new path in Fig. 2i.

**Domain wall transformations in vertices:** Local changes to the magnetic microstructure are another major factor contributing to stochastic decision-making. There are two independent sources of changes to the local magnetic microstructure, namely the use of a global applied field, and secondly the fact that the vertices at the top and bottom of each vertical bar (v1 and v2 in Fig. 1) are of different sizes. We have observed that implementation of a global field protocol had brought about changes to the local magnetic microstructure even before domain wall motion from the injection pad occurred. We recorded under-focus LTEM images of two different locations, SITE 1 and SITE 2, shown in Fig. 1, during a field-driven in situ experiment. SITE 1 corresponds to the first vertex with curved edges that is encountered by an injected domain wall, while SITE 2 is a vertex with curved edges that is further down the lattice. Typically, it is expected that upon field reversal, the domain wall is injected and moves down the honeycomb lattice while making decisions at each vertex that it encounters. Figs. 3a-d show an example of this process, in which a domain wall (ringed in yellow) moves along the direction of the applied field as it is increased from 10.5 Oe to 11.1 Oe. However, concurrent with this motion we also observe changes to the magnetic microstructure elsewhere in the nanostructure, as shown ringed in red in Figs. 3e-h.

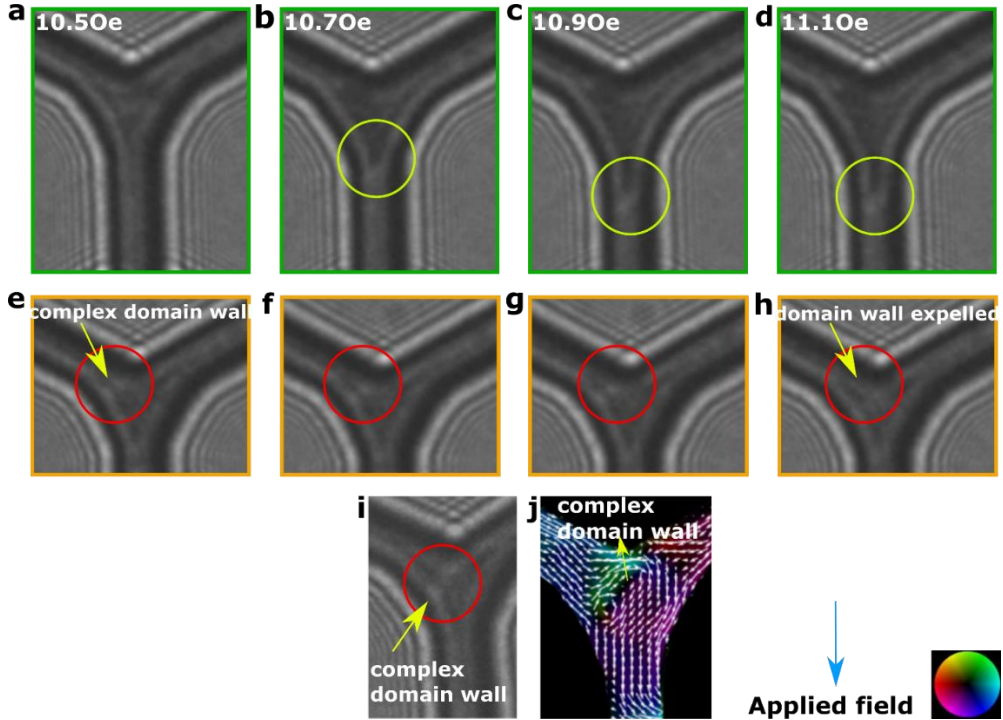

**Fig. 3**. Changes to the local magnetic microstructure. (a-d) Sequences of LTEM images showing domain wall motion at SITE 1 (see Fig. 1) leading to magnetization reversal in the adjacent bar upon application of a field. (e-h) A complex domain wall annihilation in SITE 2 (see Fig. 1) at the same time as the domain motion through SITE 1. (j) Under-focus LTEM image of a curved vertex containing a vortex-antivortex type pair. (i) Magnetic induction map of the area in (j) reconstructed using the TIE approach.

In order to understand the behavior seen in Figs. 3e-h, we need to consider the fact that the nanomagnetic Galton Boards contains two types of vertices. Each unit cell of the honeycomb lattice in the nanomagnetic Galton Boards consists of three curved vertices (at the tops of the bars) which are larger in area when compared to the vertices at the bottom of each bar, therefore the competition between exchange and dipolar energies in the larger vertices are non-trivial.[20] The increased area in the vertices with curved edges led to the emergence of complex domain walls at remanence, some which resemble a crude version of a vortex and antivortex paired in a non-trivial manner. Under global applied field conditions, some of the vortex or antivortex domain walls undergo changes to their microstructure independent of the injected domain wall. For example, the changes to the magnetic microstructure shown in Figs. 3e-h likely corresponds to annihilation of the complex vortex-antivortex wall pair. We also observed such changes to the local magnetic microstructures at locations other than the curved vertices, not shown here. We note that unfortunately, the dynamic changes to the local magnetization prevent us creating magnetic induction maps of the region shown in Figs. 3e-h. This occurs because in order to reconstruct the local magnetic microstructure, we need to solve the Transport-of-Intensity Equation (TIE) by taking a through-focal series of images at different focus conditions. Therefore, interpretation of domain wall motion is typically performed indirectly by comparing the magnetic textures observed during an in-situ experiment with those already imaged in a static condition at remanence. An example of a reconstructed in-plane magnetic induction map is provided in Fig. 3j, from which we can deduce the magnetic texture in the Y-shaped junction at remanence as shown in Fig. 3i. We then compare the magnetic induction map to that of a dynamic experiment recorded at around the same applied field condition, enabling us to argue that the change in magnetic texture observed in the dynamic experiment is likely due to annihilation of a complex structure similar to a vortex-antivortex pair. These local changes to the magnetic textures contribute to the complexity in the decision-making process, and we are currently performing additional experiments that should reveal the exact nature of these structures.

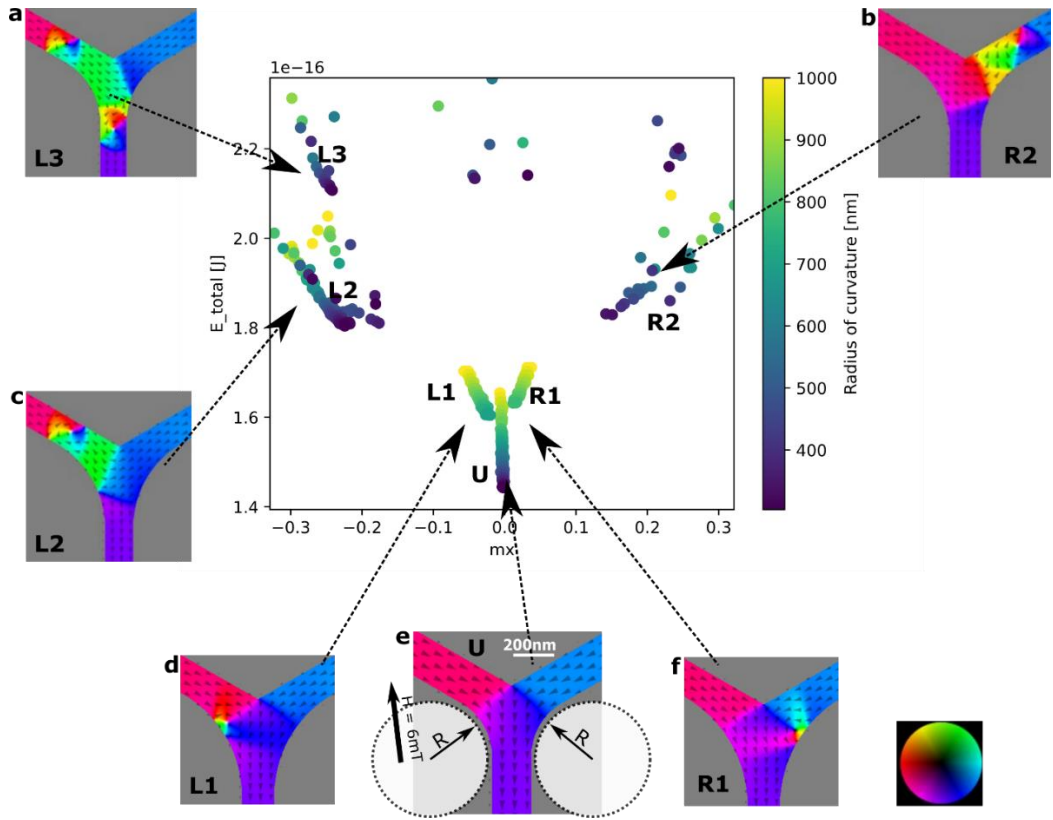

**Fig. 4**. Clustering of stable magnetic states after downwards initialization as a function of junction radius of curvature. (a-f) Simulation results based on total energy and mean value of $M_x$. Stable states are clustered according to their magnetic configuration: U, L1, L2, L3, R1, and R2. L1 and R1 correspond to states supporting a vortex core, U refers to states with uniform magnetization, and L2, L3, and R2 are higher energy states that support complex domain walls. (e) Definition of the junction radius of curvature.

To evaluate the stability of the vortex domain walls within the Y-shaped vertices, we have performed micromagnetic simulations using Mumax3. To explore all the possible stable states of the system, we initialized a vertex in a random magnetic configuration and then relaxed the system until a stable state is reached. We performed this process 15 times, employing a different initial state each time in order to sample as much as possible of the energy landscape. The resultant energetics from the simulations, along with representative schematics of the Mumax3 magnetic configurations are provided in Figs. 4a-f. In Figs. 4d-f, three different stable magnetic configurations (L1, U, and R1) are shown, in which vortex states are only stable for a radius of curvature higher than 650 nm (see Fig. 4e for a schematic of radius of curvature), whereas all the other states (L3, R2, and L2, seen in Figs. 4a-c,) are stable for the smallest simulated radius of 300 nm. We performed several simulations for different radius of curvature of the vertex in order to probe the stability of the different magnetic states as a function of vertex geometry. The end of the three arms (near the edges of the simulation) are initialized with magnetization pointing uniformly downwards and the magnetization is frozen in these end regions during the entire simulation to avoid reversing them. A uniform field of 6mT is applied upwards (antiparallel to the magnetization) to favor exploration of states other than the uniform state. An additional field of 0.6mT is applied horizontally to the left, to break the symmetry of the system and eliminate possible saddle points. To cluster the different stable states found, we mapped the simulation results into the 2D space formed by the total energy and the average values of $M_x$ (the *x* direction is horizontal in Fig. 4e), using color to distinguish simulations with a different radius of curvature. The lowest energy state is the uniform state (cluster U in Fig. 4e) and then two stable vortex states, which have a slightly higher energy (clusters L1 and R1). A higher energy state consisting of a vortex plus an edge charge is also observed (clusters L2 and R2, Figs. 4c and 4b) and finally a third cluster (L3, Fig. 4a) is observed for states consisting of two vortex domain walls, indicating an increase in energy as the wall configuration complexity increases, as expected. The increase in total energy as a function of the radius is also expected, as the total volume of magnetic material increases as the radius of curvature of the vertex is increased. The structures we have observed experimentally have a curvature of 900 nm, which is large enough for the vertices to host complex magnetic textures. Finally, the observed asymmetry of the results is expected due to the presence of a small horizontal field in the simulations. Even though it could not be proven experimentally in this work, this tilting of the energy landscape could be behind the tuneability of decision making observed previously in Galton boards upon application of transverse fields.[13]

**Conclusion**

Our work offers insights in implementing decision-making into novel computing architectures using spintronics. We have identified the sources of stochasticity through real space imaging of the domain wall motion in a field-driven nanomagnetic Galton Board, which in turn broadly relates to the physics of domain wall motion in Y-shaped junctions. We found that decision-making in the nanomagnetic Galton Boards is dependent on four factors: (1) topology of the injected domain wall leading to chiral path selectivity, as long as the applied field is below the walker breakdown, (2) changes to the domain wall's morphology brought forth by local disorder in elongated bars, (3) dissimilarly sized vertices leading to a variable contribution of local exchange and magnetostatic energies, and (4) on-set of effects related to the strength of the applied field, such as pinning and Walker Breakdown, which in combination lead to an increased complexity in the domain wall motion.

**Acknowledgements**

Work at Argonne National Laboratory (H.A and A.K.P.L.) was funded by the US Department of Energy, Office of Science, Office of Basic Energy Sciences, Materials Science and Engineering Division. Work performed at the Center for Nanoscale Materials, a U.S. Department of Energy Office of Science User Facility, was supported by the U.S.DOE, Office of Basic Energy Sciences, under Contract No.DE-AC02-06CH11357. Part of the work (D.S-H and J.G.) was supported by Quantum Materials for Energy Efficient Neuromorphic Computing (Q-MEEN-C), an Energy Frontier Research Center funded by the US Department of Energy, Office of Science, Office of Basic Energy Sciences under Award No. DE-SC0019273.

The authors would like to thank Francois Montaigne, and Daniel Lacour of Institut Jean Lamour, CNRS – Université de Lorraine, for assisting in sample fabrication.

**Conflict of Interest**

The authors declare no conflict of interest.

**Data Availability Statement**

The data supporting findings in this manuscript will be available from the corresponding author upon reasonable request.

(1) Workshop, O. o. S. Basic Research Needs for Microelectronics. Energy, D. o., Ed.; US Government: United States, 2019; p 138.
(2) Hirohata, A.; Yamada, K.; Nakatani, Y.; Prejbeanu, I.-L.; Diény, B.; Pirro, P.; Hillebrands, B. Review on spintronics: Principles and device applications. *Journal of Magnetism and Magnetic Materials* **2020**, *509*, 166711. DOI: https://doi.org/10.1016/j.jmmm.2020.166711.
(3) Grollier, J.; Querlioz, D.; Stiles, M. D. Spintronic Nanodevices for Bioinspired Computing. *Proceedings of the IEEE* **2016**, *104* (10), 2024-2039. DOI: 10.1109/JPROC.2016.2597152.
(4) Grollier, J.; Querlioz, D.; Camsari, K. Y.; Everschor-Sitte, K.; Fukami, S.; Stiles, M. D. Neuromorphic spintronics. *Nature Electronics* **2020**, *3* (7), 360-370. DOI: 10.1038/s41928-019-0360-9.
(5) Kim, Y.; Fong, X.; Roy, K. Spin-Orbit-Torque-Based Spin-Dice: A True Random-Number Generator. *IEEE Magnetics Letters* **2015**, *6*, 1-4. DOI: 10.1109/LMAG.2015.2496548.
(6) Fukushima, A.; Seki, T.; Yakushiji, K.; Kubota, H.; Imamura, H.; Yuasa, S.; Ando, K. Spin dice: A scalable truly random number generator based on spintronics. *Applied Physics Express* **2014**, *7* (8), 083001. DOI: 10.7567/apex.7.083001.
(7) Vodenicarevic, D.; Locatelli, N.; Mizrahi, A.; Friedman, J. S.; Vincent, A. F.; Romera, M.; Fukushima, A.; Yakushiji, K.; Kubota, H.; Yuasa, S.; et al. Low-Energy Truly Random Number Generation with Superparamagnetic Tunnel Junctions for Unconventional Computing. *Physical Review Applied* **2017**, *8* (5), 054045. DOI: 10.1103/PhysRevApplied.8.054045.
(8) Meier, G.; Bolte, M.; Eiselt, R.; Krüger, B.; Kim, D.-H.; Fischer, P. Direct Imaging of Stochastic Domain-Wall Motion Driven by Nanosecond Current Pulses. *Physical Review Letters* **2007**, *98* (18), 187202. DOI: 10.1103/PhysRevLett.98.187202.
(9) Arava, H.; Derlet, P. M.; Vijayakumar, J.; Cui, J.; Bingham, N. S.; Kleibert, A.; Heyderman, L. J. Computational logic with square rings of nanomagnets. *Nanotechnology* **2018**, *29* (26), 265205. DOI: 10.1088/1361-6528/aabbc3.
(10) Arava, H.; Leo, N.; Schildknecht, D.; Cui, J.; Vijayakumar, J.; Derlet, P. M.; Kleibert, A.; Heyderman, L. J. Engineering Relaxation Pathways in Building Blocks of Artificial Spin Ice for Computation. *Physical Review Applied* **2019**, *11* (5), 054086. DOI: 10.1103/PhysRevApplied.11.054086.
(11) Camsari, K. Y.; Sutton, B. M.; Datta, S. p-bits for probabilistic spin logic. *Applied Physics Reviews* **2019**, *6* (1), 011305. DOI: 10.1063/1.5055860.
(12) Sengupta, A.; Shim, Y.; Roy, K. Proposal for an All-Spin Artificial Neural Network: Emulating Neural and Synaptic Functionalities Through Domain Wall Motion in Ferromagnets. *IEEE Transactions on Biomedical Circuits and Systems* **2016**, *10* (6), 1152-1160. DOI: 10.1109/TBCAS.2016.2525823.
(13) Sanz-Hernández, D.; Massouras, M.; Reyren, N.; Rougemaille, N.; Schánilec, V.; Bouzehouane, K.; Hehn, M.; Canals, B.; Querlioz, D.; Grollier, J.; et al. Tunable Stochasticity in an Artificial Spin Network. *Advanced Materials* **2021**, *33* (17), 2008135. DOI: https://doi.org/10.1002/adma.202008135.

(14) Zeissler, K.; Walton, S. K.; Ladak, S.; Read, D. E.; Tyliszczak, T.; Cohen, L. F.; Branford, W. R. The non-random walk of chiral magnetic charge carriers in artificial spin ice. *Scientific Reports* **2013**, *3* (1), 1252. DOI: 10.1038/srep01252.

(15) Pushp, A.; Phung, T.; Rettner, C.; Hughes, B. P.; Yang, S.-H.; Thomas, L.; Parkin, S. S. P. Domain wall trajectory determined by its fractional topological edge defects. *Nature Physics* **2013**, *9* (8), 505-511. DOI: 10.1038/nphys2669.

(16) Walton, S. K.; Zeissler, K.; Burn, D. M.; Ladak, S.; Read, D. E.; Tyliszczak, T.; Cohen, L. F.; Branford, W. R. Limitations in artificial spin ice path selectivity: the challenges beyond topological control. *New Journal of Physics* **2015**, *17* (1), 013054. DOI: 10.1088/1367-2630/17/1/013054.

(17) Sethi, P.; Murapaka, C.; Goolaup, S.; Chen, Y. J.; Leong, S. H.; Lew, W. S. Direct observation of deterministic domain wall trajectory in magnetic network structures. *Scientific Reports* **2016**, *6* (1), 19027. DOI: 10.1038/srep19027.

(18) Mougin, A.; Cormier, M.; Adam, J. P.; Metaxas, P. J.; Ferré, J. Domain wall mobility, stability and Walker breakdown in magnetic nanowires. *Europhysics Letters (EPL)* **2007**, *78* (5), 57007. DOI: 10.1209/0295-5075/78/57007.

(19) Bryan, M. T.; Schrefl, T.; Allwood, D. A. Dependence of Transverse Domain Wall Dynamics on Permalloy Nanowire Dimensions. *IEEE Transactions on Magnetics* **2010**, *46* (5), 1135-1138. DOI: 10.1109/TMAG.2010.2040622.

(20) McMichael, R. D.; Donahue, M. J. Head to head domain wall structures in thin magnetic strips. *IEEE Transactions on Magnetics* **1997**, *33* (5), 4167-4169. DOI: 10.1109/20.619698.